\documentclass[letterpaper,conference]{IEEEtran}
\usepackage{amsmath,amssymb,verbatim,arydshln}
\usepackage[dvips]{graphicx}
\usepackage{setspace}

\newtheorem{definition}{Definition}[section]
\newtheorem{thm}{Theorem}[section]

\newtheorem{lemma}[thm]{Lemma}

\newtheorem{remark}{Remark}[section]

\newcommand{\tr}{{\rm Tr}}

\newcommand{\Z}{{\mathbb Z}}
\newcommand{\R}{{\mathbb R}}
\newcommand{\C}{{\mathbb C}}

\newcommand{\OO}{{\mathcal O}}

\newcommand{\Q}{{\mathbb Q}}

\begin{document}
 
\title{On the Eavesdropper's Correct Decision in Gaussian and Fading Wiretap Channels Using Lattice Codes}
\author{\IEEEauthorblockN{Anne-Maria Ernvall-Hyt\"onen$^*$ and Camilla Hollanti$^{**}$, \emph{Member, IEEE}}
\IEEEauthorblockA{$^*$Department of Mathematics and Statistics, FI-00014 University of Helsinki, Finland\\
$^{**}$Department of Mathematics,  FI-20014 University of Turku, Finland\\
}
}
\maketitle
\let\thefootnote\relax\footnotetext{The funding of A.-M. Ernvall-Hyt\"onen from the Academy of Finland (grant \#138337) is gratefully acknowledged. A.-M. Hyt\"onen would also like to thank Professor P\"ar Kurlberg  for useful discussions.
The research of C. Hollanti is supported by the Emil Aaltonen
Foundation's Young Researcher's Project, and by the Academy of Finland 
(grant \#131745). Part of this research was conducted while C. Hollanti was visiting Professor Emanuele Viterbo at the Monash University in Melbourne, Victoria, Australia. Professor Viterbo is gratefully acknowledged for the fruitful discussions during the visit that greatly influenced this paper. 
Both authors wish to thank Professors Patrick Sol\'e and Jean-Claude Belfiore for important comments.
}

%
%

\begin{abstract}
In this paper, the probability of Eve the Eavesdropper's correct decision is considered both in the Gaussian and Rayleigh fading wiretap channels when  using lattice codes for the transmission.  

First, it is proved that the secrecy function determining Eve's performance attains its maximum at $y=1$ on all known extremal even unimodular lattices. This is a special case of a conjecture by Belfiore and Sol\'e. Further, a very simple method to verify or disprove the conjecture on any given  unimodular lattice is given. 

Second, preliminary analysis on the behavior of Eve's probability of correct decision in the fast fading wiretap channel is provided. More specifically, we compute the truncated \emph{inverse norm power sum} factors in Eve's probability expression.  The analysis reveals  a performance-secrecy-complexity tradeoff: relaxing on the legitimate user's performance can significantly increase the security of transmission. The confusion experienced by the eavesdropper may be further increased by using skewed lattices, but at the cost of increased complexity.

\end{abstract}

\section{Introduction} The first part of the paper is related to the Gaussian wiretap channel \cite{Hell_Gaussianwire,belfisoleoggis,belfisole,belfioggiswire}. Belfiore and Oggier defined in \cite{belfioggiswire} the secrecy gain
\[
\max_{y\in \mathbb{R}, 0<y}\frac{\Theta_{\mathbb{Z}^n}(yi)}{\Theta_{\Lambda}(yi)},
\]
where
\[
\Theta_{\Lambda}(z)=\sum_{x\in \Lambda}e^{\pi i ||x||^2z},
\]
as a new lattice invariant to measure how much confusion the eavesdropper will experience while the lattice $\Lambda$ is used in Gaussian wiretap coding. The function $\Xi_{\Lambda}(y)=\frac{\Theta_{\mathbb{Z}^n}(yi)}{\Theta_{\Lambda}(yi)}$ is called the secrecy function. Belfiore and Sol\'e then conjectured in \cite{belfisole} that the secrecy function attains its maximum at $y=1$, which would then be the value of the secrecy gain. The secrecy gain was further studied by Oggier, Sol\'e and Belfiore in \cite{belfisoleoggis}. 

The main point of this part of the paper is to prove the following theorem:

\begin{thm}\label{main}
The secrecy function obtains its maximum at $y=1$ on all known even unimodular extremal lattices.
\end{thm}

Further, the method used here applies for any given even unimodular (and also for some unimodular but not even) lattices. This will be discussed in its own section. 

In the second part of the paper, we move from Gaussian wiretap channels on to Rayleigh fading wiretap channels.    Our attempt is to increase the understanding of the performance of wiretap lattice codes through a numerical analysis on the probability of Eve the Eavesdropper's correct decision. To this end, we provide the first explicit lattice code constructions  based on algebraic number fields $K$ and the canonical embedding of their rings of integers $\OO_K$ or an ideal $\mathcal{I}\subseteq\OO_K$, as suggested in \cite{BO_wiretap}, and  then  compute the truncated \emph{inverse norm power sum} factors in Eve's probability expression. The study concentrates on the special case of  totally real number field extensions to guarantee full diversity \cite{OV}, with three explicit example codes arising from both orthogonal and skewed lattices that are subsets in $\R^4$. The results indicate a performance-secrecy-complexity tradeoff: relaxing on the legitimate user's performance can significantly increase the security of transmission. The confusion experienced by the eavesdropper may be further increased by using skewed lattices, but at the cost of increased complexity.

We assume the fading is Rayleigh distributed. Due to lack of space, we do not repeat the channel model nor the detailed transmission scheme here, but refer to \cite{BO_wiretap} for more details. 

\section{Preliminaries and Definitions}

Let us first recall the notion of a lattice as they will play a key role throughout the paper. For our purposes, a \emph{lattice} $\Lambda$ is a discrete abelian subgroup of a real vector space, 
$$
 \Lambda=\Z \beta_1\oplus \Z \beta_2 \cdots \oplus \Z \beta_K\subset \R^n,
$$
where the elements $\beta_1,\ldots, \beta_K$ are linearly independent, \emph{i.e.}, form a lattice basis, and $K\leq n$ is
called the \emph{rank} of the lattice. Here, we only consider full rank totally real lattices, that is, we set $K=n$ and will always have $\beta_i\in \R$. 

The \emph{Gram matrix} of a lattice is  defined as 
$
G(\Lambda)=( \tr(\beta_i\beta_j^T))_{1\leq i,j\leq n}=MM^T,
$
where $M$ is the \emph{generator matrix} of the lattice. The determinant of the Gram matrix is also called \emph{lattice determinant}. The \emph{volume} of the fundamental parallelotope of the lattice is
$
\mathrm{Vol}(\Lambda)=\sqrt{\det(G(\Lambda))}=|\det(M)|.
$

\begin{definition}\label{pdmin}
The \emph{minimum product distance} of a lattice $\Lambda$  is 
$
d_{p,min}(\Lambda)=\min_{0\neq {\bf x}\in\Lambda}\prod_{i=1}^n|x_i|,
$
where ${\bf x}=(x_1,\ldots,x_n)\in\Lambda.$ 
\end{definition}
\begin{remark} In order to fairly compare different lattices, we first normalize them to a unit volume $\mathrm{Vol}(\Lambda)=1$ and then compute the (normalized) minimum product distance $d_{p,min}(\Lambda)$.
\end{remark}

A lattice is called \emph{unimodular} if its determinant $=\pm 1$, and the norms are integral, \emph{i.e.}, $||{\bf x}||^2\in \mathbb{Z}$ for all vectors ${\bf x}$ in the lattice. Further, it is called \emph{even}, if $||{\bf x}||^2$ is even. Otherwise it is called \emph{odd}. A lattice can be even unimodular only if the dimension is divisible by $8$. Odd unimodular lattices have no such restrictions.

\begin{definition}
Write $n=24m+8k$, where $k\in \{0,1,2\}$.An even unimodular lattice is called extremal if the norm of the shortest vector in the lattice is $2m+2$.
\end{definition}

It is worth noticing that the definition of extremal has changed. Earlier (see  \emph{e.g.} \cite{conwayodlyzkosloane}), extremal meant that the shortest vector was of length $\left\lfloor\frac{n}{8}\right\rfloor+1$. With the earlier definition the highest dimensional self-dual extremal lattice is in dimension $24$ (see \cite{conwayodlyzkosloane}), while with the current definition there is a selfdual extremal lattice in dimension $80$ (for a construction, see \cite{bachocnebe}).

\section{On the conjecture by Belfiore and Sol\'e in the Gaussian wiretap channel}
Let us first have a closer look at theta-functions. For an excellent source on theta-functions, see \emph{e.g.}  Chapter 10 in Stein's and Shakarchi's book \cite{steinshakarchi}.
\subsection{On theta functions}
  A theta function for a lattice $\Lambda$ is defined as follows:
\[
\sum_{x\in \Lambda} e^{-\pi i ||x||^2}.
\]

In the following, we will need functions $\vartheta_2$, $\vartheta_3$ and $\vartheta_4$, which are defined in the following way:

\begin{alignat*}{1}
\vartheta_2(\tau) & =e^{\pi i \tau/4}\prod_{n=1}^{\infty}(1-q^{2n})(1+q^{2n})(1+q^{2n-2})\\
\vartheta_3(\tau) &=\prod_{n=1}^{\infty}(1-q^{2n})(1+q^{2n-1})^2\\
\vartheta_4(\tau)  & =\prod_{n=1}^{\infty}(1-q^{2n})(1-q^{2n-1})^2.
\end{alignat*}

Let us now have a brief look at the even unimodular lattices. Write again $n=24m+8k$. Then the theta function of the lattice can be written as a polynomial of the Eisenstein series $E_4$ and the discriminant function $\Delta$:
$\Theta=E_4^{3m+k}+\sum_{j=1}^m b_i E^{3(m-j)+k}\Delta^j.$
Since $E_4=\frac{1}{2}\left(\vartheta_2^8+\vartheta_3^8+\vartheta_4^8\right)$ and $\Delta=\frac{1}{256}\vartheta_2^8\vartheta_3^8\vartheta_4^8$, the theta function of an even unimodular lattice can be easily written as a polynomial of these basic theta functions. Furthermore, since $\vartheta_2^4+\vartheta_4^4=\vartheta_3^4$, the secrecy function can be written as a simple rational function of $\frac{\vartheta_2^4\vartheta_4^4}{\vartheta_3^8}$:
\begin{multline}\label{polynomiksi}
\frac{\Theta_{\mathbb{Z}^n}}{\Theta_{\Lambda}}=\frac{\vartheta_3^{n}}{E_4^{3m+k}+\sum_{j=1}^mE_4^{3(m-j)+k}\Delta^j}\\=\left(\left(1-\frac{\vartheta_2^4\vartheta_4^4}{\vartheta_3^8}\right)^{3m+k}\right.\\ \left.+\sum_{j=1}^m\frac{b_j}{256^j}\left(1-\frac{\vartheta_2^4\vartheta_4^4}{\vartheta_3^8}\right)^{3(m-j)+k}\cdot\left(\frac{\vartheta_2^4\vartheta_4^4}{\vartheta_3^8}\right)^{2j}\right)^{-1}.
\end{multline}
Hence, finding the maximum of the secrecy function is equivalent to finding the minimum of the denominator of the previous expression in the range of $\frac{\vartheta_2^4\vartheta_4^4}{\vartheta_3^8}$.

Let us now turn to general unimodular lattices, in particular odd ones. Write now $n=8\mu+\nu$, where $n$ is the dimension of the lattice. Just like a bit earlier, the theta function of any unimodular lattice (regardless of whether it's even or odd), can be written as a polynomial (see \emph{e.g.} (3) in \cite{conwayodlyzkosloane}):
\[
\Theta_{\Lambda}=\sum_{r=0}^{\mu}a_r\vartheta_3^{n-8r}\Delta_8^r,
\]
where $\Delta_8=\frac{1}{16}\vartheta_2^4\vartheta_4^4$. Hence
\begin{equation}\label{polynomiksi2}
\frac{\Theta_{\mathbb{Z}^n}}{\Theta_{\Lambda}}=\left(\sum_{r=0}^{\mu}\frac{a_r}{16^r}\frac{\vartheta_2^{4r}\vartheta_4^{4r}}{\vartheta_3^{8r}}\right)^{-1}.
\end{equation}
Again, to determine the maximum of the function, it suffices to consider the denominator polynomial in the range of $\frac{\vartheta_2\vartheta_4}{\vartheta_3}$.

\subsection{Lemmas}

The following lemma is easy, and follows from the basic properties of theta functions. The proof will be omitted.

\begin{lemma}
Let $y\in \mathbb{R}$. The function
$f(y)=\frac{\vartheta_4^4(yi)\vartheta_2^4(yi)}{\vartheta_3^8(yi)}$
has symmetry: $f(y)=f\left({1}/{y}\right)$.
\end{lemma}

We may now formulate a lemma that is crucial in the proof of the main theorem:
\begin{lemma}\label{rajoite}
Let $y\in \mathbb{R}$. The function
$\frac{\vartheta_4^4(yi)\vartheta_2^4(yi)}{\vartheta_3^8(yi)}$
attains its maximum when $y=1$. This maximum is $\frac{1}{4}$.
\end{lemma}
\begin{proof}
To shorten the notation, write $g=e^{-\pi y}$. Notice that when $y$ increases, $g$ decreases and vice versa. Using the product representations for the functions $\vartheta_2(yi)$, $\vartheta_3(yi)$ and $\vartheta_4(yi)$, we obtain
\begin{multline*}
\frac{\vartheta_2(yi)\vartheta_4(yi)}{\vartheta_3(yi)^2}
=g^{1/4}\left(\prod_{n=1}^{\infty}(1+g^{2n})(1+g^{2n-2})\right) \\ \times \left(\prod_{n=1}^{\infty}(1-g^{2n-1})^2\right)\left(\prod_{n=1}^{\infty}(1+g^{2n-1})^{-4}\right).
\end{multline*}
Now
\[
\prod_{n=1}^{\infty}(1+g^{2n-2})=2\prod_{n=1}^{\infty}(1+g^{2n})
\]
and
\begin{multline*}
\prod_{n=1}^{\infty}(1+g^{2n-1})^{-4}=\prod_{n=1}^{\infty}(1+(-g)^n)^4\\=\left(\prod_{n=1}^{\infty}(1+g^{2n})\right)^4\left(\prod_{n=1}^{\infty}(1-g^{2n-1})\right)^4.
\end{multline*}
Combining all these pieces together, we obtain
\[
\frac{\vartheta_2(yi)\vartheta_4(yi)}{\vartheta_3(yi)^2}=
2\left(g^{1/24}\prod_{n=1}^{\infty}(1+(-g)^n)\right)^6.
\]
Since the factor $2$ is just a constant, it suffices to consider the function $g^{1/24}\prod_{n=1}^{\infty}(1+(-g)^n)$. To find the maximum, let us first differentiate the function:
\begin{multline*}
\frac{\partial}{\partial g}\left(g^{1/24}\prod_{n=1}^{\infty}(1+(-g)^n)\right)\\=\left(g^{1/24}\prod_{n=1}^{\infty}(1+(-g)^n)\right)\left(\frac{1}{24g}+\sum_{n=1}^{\infty}\frac{n(-1)^ng^{n-1}}{1+(-g)^n}\right).
\end{multline*}
Since $g^{1/24}\prod_{n=1}^{\infty}(1+(-g)^n)$ is always positive, it suffices to analyze the part $\frac{1}{24g}+\sum_{n=1}^{\infty}\frac{n(-1)^ng^(n-1)}{1+(-g)^n}$ to find the maxima. We wish to prove that the derivate has only one zero, because if it has only one zero, then this zero has to be located at $y=1$ (because the original function has symmetry, and therefore, a zero in the point $y$ results in a zero in the point $\frac{1}{y}$ which has to be separate unless $y=1$). To show that the derivative has only one zero, let us consider the second derivative, or actually, the derivative of the part $\frac{1}{24g}+\sum_{n=1}^{\infty}\frac{n(-1)^ng^{n-1}}{1+(-g)^n}$. Now
\begin{multline*}
\frac{\partial}{\partial g}\left(\frac{1}{24g}+\sum_{n=1}^{\infty}\frac{n(-1)^ng^{n-1}}{1+(-g)^n}\right)\\=-\frac{1}{24g^2}+\sum_{n=1}^{\infty}\left(\frac{n(n-1)(-1)^ng^{n-2}}{1+(-g)^n}-\frac{n^2g^{2(n-1)}}{(1+(-g)^n)^2}\right).
\end{multline*}
Now we wish to show that this is negative when $g\in(0,1)$. Let us first look at the term $-\frac{1}{24g^2}$ and the terms in the sum corresponding the values $n=1$ and $n=2$. Their sum is
\begin{multline*}
-\frac{1}{24g^2}-\frac{1}{(1-g)^2}+\frac{2-2g^2}{(1+g^2)^2}\\=\frac{-73g^6+98g^5-51g^4-92g^3+21g^2+2g-1}{24g^2(1-g)^2(1+g^2)^2}.
\end{multline*}
The denominator is positive when $g\in (0,1)$, and the nominator has two real roots, which are both negative (approximately $g_1\approx-0.719566$ and $g_2\approx-0.196021$). On positive values of $g$, the nominator is always negative. In particular, the nominator is negative when $g\in (0,1)$.

Let us now consider the terms $n>2$, and show that the sum is negative. Since the original function has symmetry $y\rightarrow \frac{1}{y}$, and we are only considering the real values of the theta series, we may now limit ourselves to the interval $y\in [1,\infty)$, which means that $g\in (0,e^{-\pi}]$.  Let us now show that the sum of two consecutive terms where the first one corresponds an odd value of $n$, and the second one an even value of $n$ is negative. The sum looks like the following:
\[
-\frac{n(n-1)g^{n-2}}{1-g^n}-\frac{n^2g^{2(n-1)}}{(1-g^n)^2}+\frac{n(n+1)g^{n-1}}{1+g^{n+1}}-\frac{(n+1)^2g^{2n}}{(1+g^{n+1})^2}.
\]
Let us estimate this, and take a common factor:
\begin{multline*}
<g^{n-2}n\left(-\frac{n-1}{1-g^n}-\frac{ng^n}{(1-g^n)^2}\right.\\ \left.+\frac{(n+1)g}{1+g^{n+1}}-\frac{(n+1)g^{n+2}}{(1+g^{n+1})^2}\right)\\=g^{n-2}n\left(-\frac{n-1+g^n}{(1-g^n)^2}+\frac{(n+1)g}{(1+g^{n+1})^2}\right)\\<g^{n-2}n\left(\frac{-(n-1)-g^n+(n+1)g}{(1+g^{n+1})^2}\right)<0,
\end{multline*}
when
$(n-1)+g^n>(n+1)g.$
Since $(n-1)+g^n>(n-1)$, and $(n+1)g\leq (n+1)e^{-\pi}<\frac{n+1}{10}<n-1$, when $n\geq 2$, this proves that the first derivative has only one zero. This zero is at $y=1$. Since the second derivative is negative, it means that this point is actually the maximum of the function. The maximum value is
$\frac{\vartheta_2^4(i)\vartheta_4^4(i)}{\vartheta_3^8(i)}=\frac{1}{4}$.
\end{proof}

\subsection{Proof of the main theorem}
Let us first deal with  the lattice $E_8$ as a warm-up case.
We wish to show that
\begin{thm}\label{tavoite}
$\Xi_{E_8}(y)\leq \Xi_{E_8}(1).$
\end{thm}
\begin{proof}
 Notice that
\begin{multline*}
\Xi_{E_8}(y)=\left(\frac{1}{2}\left(\frac{\vartheta_2(yi)^8+\vartheta_3(yi)^8+\vartheta_4(yi)^8}{\vartheta_3(yi)^8}\right)\right)^{-1}\\
=\left(1-\frac{\vartheta_2^4(yi)\vartheta_4^4(yi)}{\vartheta_3^8(yi)}\right)^{-1},
\end{multline*}
Therefore, to show that Theorem \ref{tavoite} holds, it suffices to show that
$\frac{\vartheta_2(yi)^4\vartheta_4(yi)^4}{\vartheta_3(yi)^8}\leq \frac{\vartheta_2(i)^4\vartheta_4(i)^4}{\vartheta_3(i)^8}$,
which is equivalent to showing that 
$\frac{\vartheta_2(yi)\vartheta_4(yi)}{\vartheta_3(yi)^2}\leq \frac{\vartheta_2(i)\vartheta_4(i)}{\vartheta_3(i)^2},$
which we have already done in Lemma \ref{rajoite}.
\end{proof}

Let us now concentrate on the other cases. Again, write $z=\frac{\vartheta_2^4\vartheta_4^4}{\vartheta_3^8}$. The following table gives the secrecy functions of all known extremal even unimodular lattices (notice that these are known only in dimensions $8-80$):\\[0.1cm]
{\tiny \begin{tabular}{c|c}
dimension & $\Xi$ \\ \hline
$8$ & $\left(1-z\right)^{-1}$\\
$16$ & $\left((1-z)^{2}\right)^{-1}$ \\
$24$ & $\left((1-z)^3-\frac{45}{16}z^2\right)^{-1}$ \\
$32$ & $\left((1-z)^4-\frac{15}{4}(1-z)z^2\right)^{-1}$ \\
$40$ & $\left((1-z)^5-\frac{75}{16}(1-z)^2z^2\right)^{-1}$ \\
$48$ & $\left((1-z)^6-\frac{45}{8}(1-z)^3z^2+\frac{3915}{2048}z^4\right)^{-1}$\\
$56$ & $\left((1-z)^7-\frac{105}{16}(1-z)^4z^2+\frac{21735}{4096}(1-z)z^4\right)^{-1}$ \\
$64$ & $\left((1-z)^8-\frac{15}{2}(1-z)^5z^2+\frac{4905}{512}(1-z)^2z^4\right)^{-1}$ \\
$72$ & $\left((1-z)^9-\frac{135}{16}(1-z)^6z^2+\frac{60345}{4096}(1-z)^3z^4-\frac{53325}{32768}z^6\right)^{-1}$\\
$80$ & $\left((1-z)^{10}-\frac{75}{8}(1-z)^7z^2+\frac{42525}{2048}(1-z)^4z^4-\frac{202125}{32768}(1-z)z^6\right)^{-1}$
\end{tabular}}\\[0.3cm]
It suffices to show that the first derivatives of the denominators are negative because then the denominator is decreasing, and the function is increasing and obtains its maximum at $z=\frac{1}{4}$. It is a straightforward calculation to show this. For some details, see \cite{amenarxivepaperi}.

\subsection{Method for any given unimodular lattice}
Let $\Lambda$ be a  unimodular lattice. Then its secrecy function can be written as a polynomial $P(z)$, where $z=\frac{\vartheta_2^4\vartheta_4^4}{\vartheta_3^8}$ as shown in (\ref{polynomiksi}) and \eqref{polynomiksi2}. Now, according to Lemma \ref{rajoite}, $0\leq z\leq \frac{1}{4}$ (the lower bound does not follow from the lemma but from the fact that $z$ is a square of a real number). Therefore, it suffices to consider the polynomial $P(z)$ on the interval $[0,\frac{1}{4}]$. The conjecture is true if and only if the polynomial obtains its smallest value on the interval at $\frac{1}{4}$. Investigating the behaviour of a given polynomial to show whether one point is its minimum on a short interval is a very straightforward operation.

%
%

\section{On the size of Eve's inverse norm power sum in a fast Rayleigh fading wiretap channel}
Let us now look at the Rayleigh fading wiretap channel and analyze the behavior of the probability for Eve's correct decision in some example cases. This will give us a preliminary understanding as to what are the key properties affecting the secrecy gained by lattice coding. 
\subsection{The probability expression and the inverse norm power sum}
We start by  recalling the expression $P_{c,e}$ for the probability of a correct decision for Eve, when  observing a  lattice  $ \Lambda_e$.
 For the fast fading case \cite[Sec.III-A]{BO_wiretap},
\begin{equation}\label{fast-prob}
P_{c,e}\simeq \left(\frac 1{4\gamma_e^2}\right)^{n/2}\textrm{Vol}(\Lambda_b) \sum_{0\neq\mathbf{x}\in\Lambda_e}\prod_{i=1}^n\frac{1}{|x_i|^3},
\end{equation}
where $\gamma_e$ is the average SNR for Eve assumed sufficiently large so that Eve can perfectly decode $\Lambda_e$. Here $\Lambda_b$ denotes the lattice intended for Bob, and $\Lambda_e\subset \Lambda_b$. It  can be concluded from \eqref{fast-prob} that the smaller the sum
$
\sum_{0\neq\mathbf{x}\in\Lambda_e}\prod_{i=1}^n\frac{1}{|x_i|^3},
$
the more confusion  Eve is experiencing.

 As a construction method, the authors of \cite{BO_wiretap} propose to use the canonical embedding of the ring of integers $\OO_K$ (or a suitable proper ideal $\mathcal{I}\subset\OO_K$) of a number field  $K$ over $\Q$. The field $K$ is chosen totally real to achieve full diversity. More precisely,  if $x\in\OO_K$, the transmitted lattice vector in the fast fading case would be 
\begin{equation}\label{embvector}
\mathbf{x}=(\sigma_1(x),\sigma_2(x),\ldots,\sigma_{n}(x))\in\OO_K^n=\Lambda_e,
\end{equation}
where $\sigma_i$ are the (now all real) embeddings of $K$ into $\C$. The corresponding probability for Eve's correct decision \eqref{fast-prob} yields the following \emph{inverse norm power sum} to be minimized \cite[Sec.III-B]{BO_wiretap}:
\begin{equation}\label{fast-sum}
S_M=\sum_{x\in\OO_K}\frac{1}{|N_{K/\Q}(x)|^3},
\end{equation}
where $M$ denotes the generator matrix of the lattice $\Lambda_e$.

\begin{remark}
The infinite sums above do not necessarily converge. In practice, however, the sum will always be truncated as $\mathbf{x}\in \mathcal{C}\subsetneq \Lambda_e$, where the code $\mathcal{C}$ is finite.
\end{remark}

\subsection{Example constructions and analysis on the sum $S_M$}

In this section, we describe three alternative constructions for the fast fading channel built from different number fields and their ideals. Optimal  and nearly optimal unitary lattice generator matrices in terms of the minimum product distance (cf. Def. \ref{pdmin}) are provided in \cite{Viterbo_rotations}. We will analyze  the ones with degree $n=4$, denoted here by $\Lambda_{1}$  and $\Lambda_{2}$, with the   respective unitary (\emph{i.e.}, $MM^T=I_4$) generator matrices 
$M_{1}$ (\cite[optimal, $M_1=\mathrm{krus\_4}$ ]{Viterbo_rotations})
and 
$M_{2}$ (\cite[suboptimal, $M_2=\mathrm{mixed\_2x2}$]{Viterbo_rotations}). The first construction is based on the Kronecker product of the lattice generator matrices corresponding to the canonical embeddings of the rotated $\Z^2$ lattices  $\alpha_1\Z[\sqrt{2}]$ and $\alpha_2\Z[\theta]$, where   $ \theta=\frac{1+\sqrt5}2,\,\alpha_1=\frac{1}{2\sqrt{2}+4}$ and $\alpha_2=3-\theta$. The second construction corresponds to the canonical embedding of  $\OO_{\Q(\delta)}$, where $\delta^4-\delta^3-3\delta^2+\delta+1=0$. Both lattices are rotated versions of $\Z^4$ with full diversity and good minimum product distances, $d_{p,min}(\Lambda_{1})=\frac1{\sqrt{5^2\cdot 29}}\approx 0.037139...
$ and $d_{p,min}(\Lambda_{1})=\frac1{40}\approx 0.025$. We use finite constellations $\mathcal{S}_m$  constructed by taking a square box with a zero mean within the lattice, \emph{i.e.},
 $$
 \mathbf{x}\in\mathcal{S}_m=\left\{\sum_{i=1}^n z_ix_i\, \bigg|\, m\geq z_i\in\Z\right\}\subset \Lambda_e.
$$ 
Let us now compare these two (finite) orthogonal constructions by computing truncated sums 
\begin{equation}
S_{M}(P_{lim},m)=\sum_{0\neq \mathbf{x}\in\Lambda_e\cap \mathcal{S}_m,||\mathbf{x}||_E^2\leq P_{lim}}\frac{1}{|N_{K/\Q}(x)|^3}
\end{equation}
 for a given power limit $P_{lim}$ and for a given finite constellation $\mathcal{S}_m$. In the above sum, $\mathbf{x}=(x_1,\ldots,x_n)=(\sigma_1(x),\sigma_2(x),\ldots,\sigma_{n}(x))$, where $x\in\OO_K$ or $x\in \mathcal{I}\subset\OO_K$. For a fair comparison, the lattices are normalized to unit energy,  \emph{i.e.}, to have $\mathrm{Vol}(\Lambda_e)=1$. The volumes of the corresponding superlattices $\Lambda_b$ of Bob will then scale accordingly. 

In Table I we have listed the inverse norm power sums for fixed constellations without limiting the energy, that is, the codebook will be of size $|\mathcal{C}_{ort}|=(2m+1)^4$. The maximum energies $P_{max}$ used by the constellations are also provided. 

From Table I we can make the following important conclusion. In terms of the pair-wise error probability (PEP)  for Bob as the intended legitimate receiver, the optimal lattice is known to provide (at least asymptotically) the best performance. However, from the secrecy point of view the suboptimal lattice may provide significantly improved secrecy by causing more confusion to the eavesdropper Eve. This is due to a secondary code design criterion related to maximizing the norms of the code vectors (usually showing its PEP effect at the low SNR regime), which obviously plays an important role also in the wiretap scenario (cf. \eqref{fast-sum}).

Next, we extend our analysis by computing the inverse norm power sums for a skewed lattice, denoted by $\Lambda_{3}$, corresponding to the maximal real subfield of the 15th cyclotomic field. The generator matrix is denoted by $M_3$. The minimum product distance of this lattice is $d_{p,min}(\Lambda_{3})=\frac1{\sqrt{1125}}\approx 0.02981...$ putting it in between the lattices $\Lambda_{1}$ and $\Lambda_{2}$ in terms of $d_{p,min}(\Lambda)$. From Table II, we can conclude that skewed lattices may significantly increase the secrecy compared to orthogonal lattices. One has to notice, however, that this bares the price of increased complexity as we need to carve spherical codebooks by using a bigger alphabet in order to get the possible benefits.  More precisely, we only choose the codewords in the set $\{\mathbf{x}\in\Lambda_e\cap \mathcal{S}_m\,|\, ||x||_E^2\leq P_{lim}\}.$ Hence, in order to achieve the same size of a codebook that we would have without an energy limit, we may need to  increase $m$ (see \emph{e.g.} the boldface lines in Table II). The bigger the $m$, the closer  we get to a spherical constellation with a given energy limit.

\begin{table}[h!]\label{table1}
\caption{Values of $S_{M}(P_{lim}=\infty,m)$ for orthogonal lattices without an additional energy limit and with a codebook size  $|\mathcal{C}_{ort}|=(2m+1)^4$.}
\begin{center}
{\footnotesize \begin{tabular}{|r|r|r|l|c|}\hline
 $m$ &$P_{max}$&$P_{ave}$& $S_{M_1}(P_{lim},m)$ & $S_{M_2}(P_{lim},m)$\\
 \hline
 $1$& $ 4$& $2.67$&  $9.12264\cdot 10^7$&$2.83706\cdot 10^6$  \\$
 2$& $ 16$ & $8.00$& $2.24565\cdot 10^{10}$ &$6.46037\cdot 10^6$ \\$
 \bf{3}$& $ {\bf 36}$& ${\bf 16.00}$&${\bf 2.49382\cdot 10^{11}}$ & $\mathbf{1.16395\cdot 10^7}$ \\$
 4$& $ 64$& $26.67 $& $2.49829\cdot 10^{11}$&$1.52838\cdot 10^7$ \\$
 5 $ & $ 100 $ & $40.00  $ & $2.49851\cdot 10^{11}$& $1.99487\cdot 10^7$ \\$
 6 $ & $ 144 $ & $56.00 $ & $2.50437\cdot 10^{11}$ &$2.38188\cdot 10^7$ \\$ 
 7 $ & $ 196 $ & $74.67  $ & $2.61395\cdot 10^{11}$&$2.69652\cdot 10^7$  \\$
8 $ & $ 256 $ & $96.00  $ & $2.61736\cdot 10^{11}$&$3.00791\cdot 10^7$ \\$
  9 $ & $ 324 $ & $120.00 $ & $2.61739\cdot 10^{11}$ &$3.42272\cdot 10^7$ \\$
   10  $ & $ 400 $ & $146.67 $ &  $2.71764\cdot 10^{11}$& $3.68287\cdot 10^7$ \\
\hline
\end{tabular}}
\end{center}
\end{table}
\begin{table}[h!]\label{table1}
\caption{Values of $S_{M}(P_{lim},m)$ for a skewed lattice with bounded energy.}\vspace{-0.4cm}
\begin{center}
{\scriptsize \begin{tabular}{|r|r|r|r|r|r|r|}\hline
 $m$ &$P_{lim}$&$P_{max}$&$P_{ave}$& $|\mathcal{C}_{sph}|$&$|\mathcal{C}_{ort}|$& $S_{M_3}(P_{lim},m)$ \\
 \hline
$ 8 $ & $ 4 $ & $3.63 $ & $2.66 $ &  $79$&$81$&$1.89195\cdot 10^6$  \\$
 5 $ & $ 16 $ & $15.71  $ & $9.18 $ &  $555$ &$625$&$4.24298\cdot 10^6$ \\$
 6 $ & $ 16 $ & $15.71 $ & $9.56 $ & $715$   & $625$&$4.77423\cdot 10^6$ \\$
 {\bf 7}$& ${\bf 36}$&${\bf 35.57}$&$\bf{20.33}$& ${\bf 2405}$&${\bf 2401}$&${\bf 7.13024\cdot 10^6}$ \\$
{\bf 12}$&${\bf 36}$&${\bf 24.00}$&${\bf 15.24}$&${\bf 2401}$&${\bf 2401}$&${\bf 2.29374\cdot 10^6}$\\$
 9$& $64$&$63.89$&$35.67$& $6929$&$6561$& $9.93903\cdot 10^6$ \\$
 10$& $100$&$99.97$&$55.72$&$13663$ &$14641$&$1.20680\cdot 10^7$ \\$
 11$& $100$&$99.97$&$55.57$& $16053$&$14641$&$1.29038\cdot 10^7$  \\$
  14$& $196$&$195.98$&$106.63$& $50975$&$50625$&$1.29038\cdot 10^7$  \\$
  18$& $324$&$323.93$& $175.95$ &$137273$&$130321$&$2.18703\cdot 10^7$ \\$
  20$&$400$&$399.90$&$217.31$&$208411$&$194481$&$2.40716\cdot 10^7$\\
 \hline
\end{tabular}}
\end{center}
\end{table}

Note that we have normalized the lattices to a unit volume (corresponding to a unit minimum energy in the orthogonal case), whereas to compare the full probability expressions \eqref{fast-prob} we should  normalize the SNR term rather with respect to a unit average energy. For comparison purposes, this makes no difference for orthogonal lattices as the average energies are directly determined  by the signaling alphabet and not affected by the generator matrices, so the scaling factors will coincide. However, in the case of skewed lattices the situation is different, and the average energy has an input coming from the generator matrix in addition to the alphabet. This may loosen our conclusion related to skewed lattices to some extend. Due to lack of time, we studied this effect here only for the case of maximum energy/energy limit 36 (see the boldface lines in Table I and Table II).  We can see that the skewed lattice can achieve even better energy distribution than the orthogonal ones, when $m$ is chosen sufficiently large. Unfortunately, the bigger the $m$, the higher the complexity. Further analysis is clearly required and planned to be carried out by the final submission of this paper. Due to lack of space, we were also forced to omit the corresponding analysis on the block fading channel. This will be reported in near future.

%
%

\section{Conclusions}
\label{sec:conc}

We analyzed the probability of Eve the Eavesdropper's correct decision in a wiretap channel. In the case of a Gaussian wiretap channel, we proved the Belfiore-Sol\'e conjecture for all known even unimodular extremal lattices, and gave a method to prove or disprove it on any given unimodular lattice. In the case of a Rayleigh fading wiretap channel, we computed truncated values of the inverse norm power sum and compared three different lattices. The comparison resulted in an interesting conclusion: slightly relaxing on Bob's optimal  performance can significantly increase the secrecy of the transmission.  Further reliability can be achieved by using a skewed lattice, but at the cost of increased complexity. Hence, there is clearly a performance-secrecy-complexity tradeoff aiding the service provider's subjective choice for  a suitable transmission scheme.

%
%

\end{document}